\documentclass[a4paper]{article}

\usepackage{INTERSPEECH2018}
\usepackage{subfiles}
\usepackage{color}
\usepackage{amsmath}
\usepackage{balance}
\title{Learning Spontaneity to Improve Emotion Recognition in Speech}
\name{Karttikeya Mangalam, Tanaya Guha}

\DeclareMathOperator*{\argmin}{arg\,min}

\address{
  Indian Institute of Technology, Kanpur}
\email{mangalam@iitk.ac.in, tanaya@iitk.ac.in}

\begin{document}

\maketitle
\begin{abstract}
We investigate the effect and usefulness of spontaneity (i.e.\ whether a given speech is spontaneous or not) in speech in the context of emotion recognition. We hypothesize that emotional content in speech is interrelated with its spontaneity, and use spontaneity classification as an auxiliary task to the problem of emotion recognition. We propose two supervised learning settings that utilize spontaneity to improve speech emotion recognition: a hierarchical model that performs spontaneity detection before performing emotion recognition, and a multitask learning model that jointly learns to recognize both spontaneity and emotion. Through various experiments on the well-known IEMOCAP database, we show that by using spontaneity detection as an additional task, significant improvement can be achieved over emotion recognition systems that are unaware of spontaneity. We achieve state-of-the-art emotion recognition accuracy (4-class, 69.1\%) on the IEMOCAP database outperforming several relevant and competitive baselines.
\end{abstract}

\section{Introduction}
\label{sec:intro}
Recognizing human emotion is critical for any human-centric system involving human-human or human-machine interaction. Emotion is expressed and perceived through various verbal and non-verbal cues, such as speech and facial expressions. In the recent years, speech emotion recognition has been studied extensively, both as an independent modality \cite{speechalone1,abdelwahab2017ensemble}, and in combination with others \cite{multi1}.  

The majority of work on speech emotion recognition follows a two step approach. First, a set of acoustic and prosodic features are extracted, and then a machine learning system is employed to recognize the emotion labels \cite{jin2015speech, abdelwahab2017ensemble, zong2016cross, nwe2003speech, schuller2003hidden}. Although acoustic and prosodic features are more common, the use of lexical features, such as emotional vector, have been also shown to be useful \cite{jin2015speech}. For recognition, various methods have been proposed - starting from traditional hidden Markov models (HMM) \cite{schuller2003hidden} to ensemble classifiers \cite{lee2005toward}, and more recently, deep neural networks \cite{lim2016speech, neumann2017attentive, emointerspeech}. Recently, Abdelwahab and Busso \cite{abdelwahab2017ensemble} proposed an ensemble feature selection method that addresses the problem of training and test data arising from different distributions. Zong et al.~\cite{zong2016cross} introduced a domain-adaptive least squares regression technique for the same problem. Owing to the latest trends in machine learning, autoencoders \cite{deng2014autoencoder} and recurrent neural networks (RNN) \cite{lim2016speech} have also been used for speech emotion recognition. Efforts to improve speech emotion recognition are primarily concentrated on building a better machine learning system. Although spontaneity, fluency and nativity of speech are well studied in the literature, their effect on emotion recognition tasks is not well studied. Work that addressed the problem of distinguishing between spontaneous and scripted speech include acoustic and prosodic feature-based classification \cite{spo1, spo2}, and detecting target phonemes \cite{spo3}. Dufour et al.~\cite{spo2} have also shown that spontaneity is useful for identifying speakers' role \cite{spo2} by utilizing spontaneity information in their automatic speech recognition system. A recent work by Tian et al.\ \cite{tian2015emotion} has established that emotional content is essentially different in spontaneous vs.\ acted speech (prepared, planned, scripted). As they compare emotion recognition in the two types of speech, they observe that different sets of features contribute to the success of emotion classification in spontaneous vs.\ acted speech. Another study on emotion recognition using convolutional neural network (CNN) has found that type of data (spontaneous or not) does affect the emotion recognition results \cite{neumann2017attentive}; however, this work does not use spontaneity information in emotion recognition task.  A very recent work has used gender and spontaneity information explicitely in a long short term memory (LSTM) network for effective speech emotion recognition in an aggregated data corpus \cite{emointerspeech}. Our work differs from this work by providing a detailed analysis and insight towards the effect of spontaneity in emotion recognition in speech, and by proposing an SVM-based hierarchical and multitask learning framework.

In this work, we investigate the usefulness of spontaneity in speech in the context of emotion recognition. We hypothesize that emotional content is interrelated with the spontaneity of speech, and propose to use spontaneity classification as an auxiliary task to the problem of emotion recognition in speech. We investigate two supervised learning settings: (i) a \emph{multilabel hierarchical model} that performs spontaneity detection followed by emotion classification, and (ii) a \emph{multitask learning model} that jointly learns to recognize both spontaneity and emotion in speech and returns two labels. To construct the proposed models, we use a set of standard acoustic and prosodic features in conjunction with support vector machine (SVM) classifiers. We choose SVM because it has been shown to produce results comparable to long short term memory (LSTM) networks when the training dataset is not sufficiently large \cite{tian2015emotion}. Through experiments on the IEMOCAP database \cite{iemocap}, we observe that (i) recognizing emotion is easier in spontaneous speech than in scripted speech, (ii) longer context is useful in spontaneity classification, and (iii) significant improvement in emotion recognition can be achieved using spontaneity as an additional information, over spontaneity-unaware systems.

The rest of this paper is organized as follows: Section 2 describes the feature extraction process and the two supervised classification methods we have used for using spontaneity in emotion classification. Section 3 provides details on the experimental setup and results, followed by conclusion in Section 4.
%--------------------------------------
\section{Emotion Recognition using Spontaneity}
\label{sec:pagestyle}
In this section, we propose two models that utilize the spontaneity information in speech to improve emotion recognition: (i) a multilabel hierarchical model that performs spontaneity detection followed by emotion recognition, and (ii) a multitask learning model that jointly recognizes both spontaneity and emotion labels. 
\subsection{Feature extraction}
\label{subsec:features}
We extract a set of speech features following the Interspeech2009 emotion challenge \cite{isemo}. The feature set includes four low level descriptors (LLDs) - Mel-frequency cepstral coefficients (MFCC), zero-crossing rate (ZCR), voice probability (VP) computed using autocorrelation function, and fundamental frequency (F0). For each speech sample, we use a sliding window of length $w$ with a stride length $m$ to extract the LLDs. This generates a $k$ dimensional local feature vector for each windowed segment. Each descriptor is then smoothed using a moving average filter, and the smoothed version is used to compute their respective first order delta coefficients. Appending the delta features, we obtain a local feature vector of dimension $2k$ for every windowed segment. To create a global feature for the entire speech sample, the local features are pooled temporally by computing $12$ different statistics (e.g.~mean, range, max, kurtosis) along each of the $2k$ dimensions, generating a global feature vector $\mathbf{f}\in\mathbb{R}^{d}$, $d = 24k$ for each data sample. 
\subsection{Multilabel hierarchical emotion recognition}
\label{subsec:heirarchical}
Let us consider $N$ training samples and their corresponding feature representations $\mathbf{F} = \{\mathbf{f}_j\}_{i=1}^N$, where $\mathbf{f}_j\in \mathbb{R}^d$. Each training sample with feature vector $\mathbf{f}_j$ is associated with two labels $y_j = \{y_j^s, y_j^e \}$, where $y_j^s\in Y^s$, $Y^s = {0,1}$ represents the binary spontaneity labels, and $y_j^e\in Y^e$, $Y^e= {0,1,2,3}$ denotes the emotion labels. Note that only four emotion labels are considered in this paper. We denote the entire label space as $Y = Y^s \times Y^i$.
\begin{figure}[tb]
\centering
\includegraphics[width=1.0\linewidth,trim={1cm 2cm 0cm 2cm}]{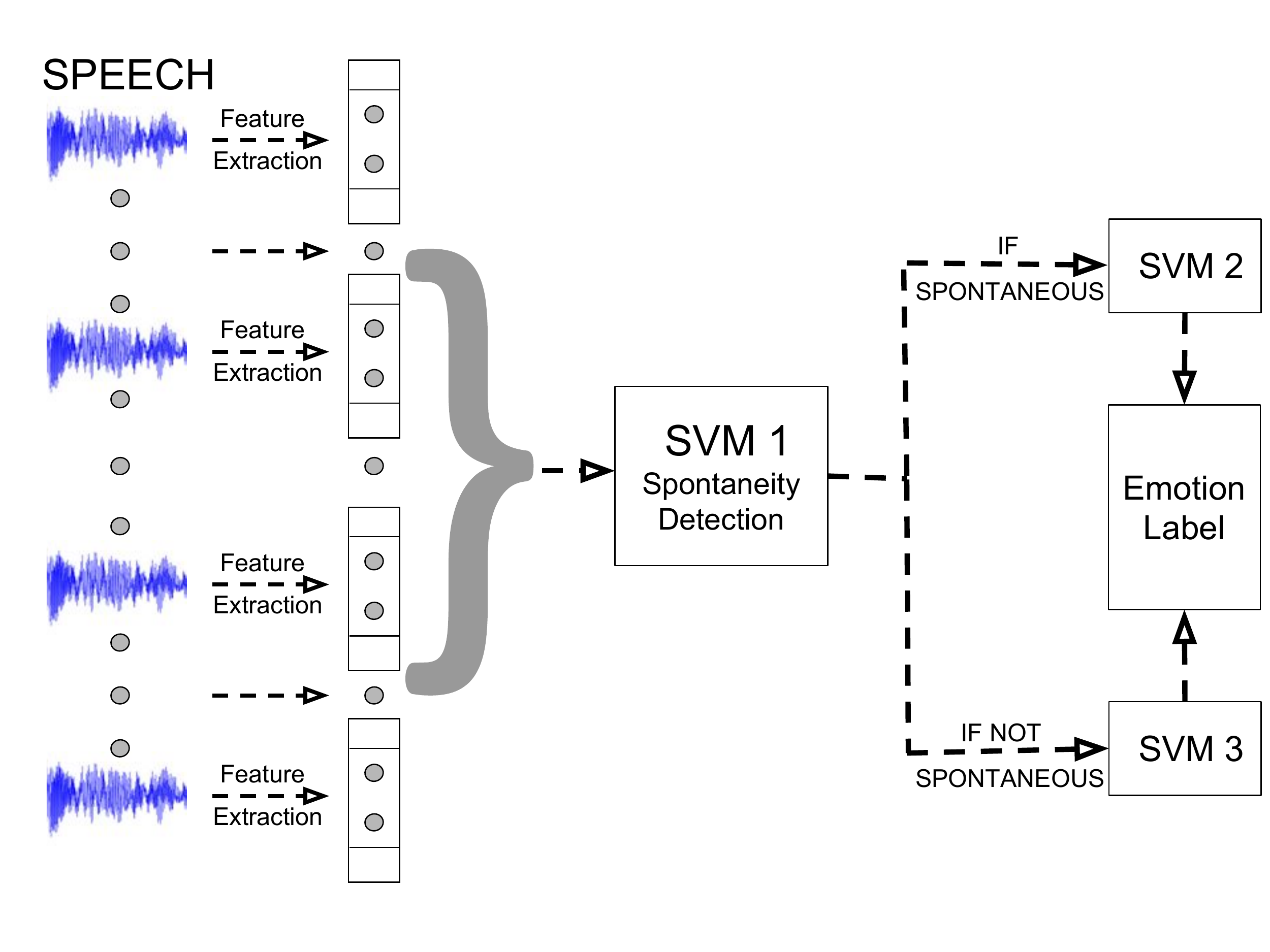}
\caption{Multilabel hierarchical emotion recognition using spontaneity}
\label{fig:m1}
\end{figure}
\par
In order to use the spontaneity information in speech, we propose a simple system which first recognizes if a speech sample is spontaneous or not. An emotion classifier is then chosen based on the decision made by the spontaneity classifier. We divide the entire training set $\Omega_{train}$ of $N$ samples into two subsets: $\Omega^1_{train}$ that contains all the spontaneous speech samples, and $\Omega^0_{train}$ that contains all the scripted or the non-spontaneous samples. As shown in Figure~\ref{fig:m1}, we train two separate support vector machine (SVM) classifiers for recognizing emotion using $\Omega^0_{train}$ and $\Omega^1_{train}$. Additionally, we train another SVM for spontaneity detection with sequence length $\ell$ (denotes the number of consecutive utterances in an input sample) using $\mathbf{F}$ on entire $\Omega_{train}$. The sequence length $\ell$ is used to account for the context needed to recognize spontaneity, which is known to help in emotion recognition \cite{gupta2014multimodal}. Later, in Section \ref{subsec:impspontaneity}, we investigate the role of $\ell$ in spontaneity detection. Note that only the spontaneity classifier uses a sequence length of $\ell=10$, but the emotion recognition is performed at utterance level.
\subsection{Multitask learning for emotion and spontaneity}
\label{subsec:mtl}
\begin{figure}[tb]
\centering
\includegraphics[width=1.0\linewidth,trim={1cm 6cm 0cm 6cm}]{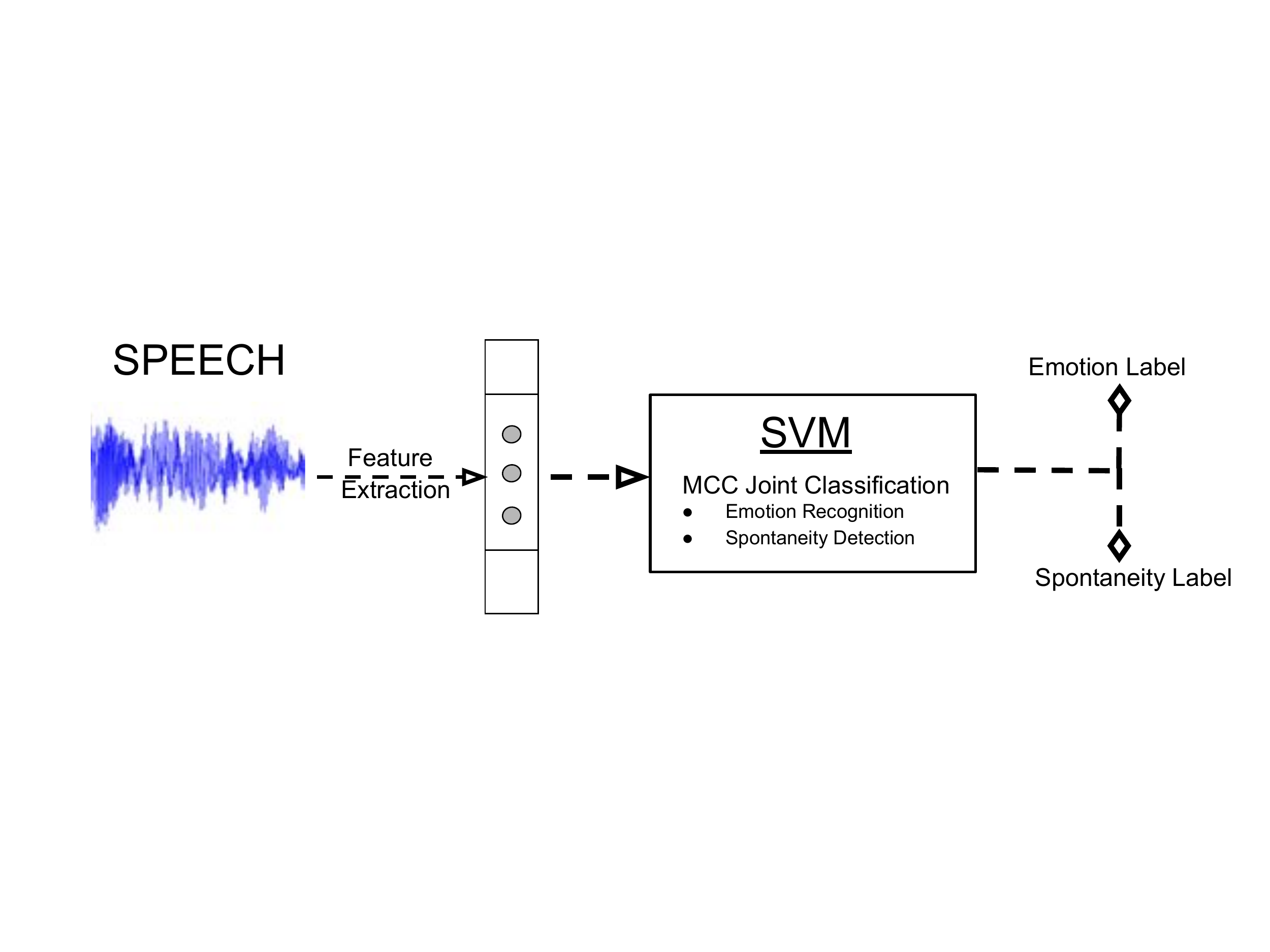}
\caption{Joint emotion and spontaneity classification.} 
\label{fig:m2}
\end{figure}
According to our hypothesis, spontaneity and emotional information in speech are interrelated. We perform the tasks of spontaneity detection and emotion recognition together in a multitask learning framework. Instead of focusing on a single learning task, a multitask learning paradigm shares representations among related tasks by learning simultaneously, and enables better generalization \cite{mtlcaruna}. Following this idea, we jointly learn to classify both spontaneity and emotion. This is posed as a multilabel multioutput classification problem. The basic idea is presented in Fig.~\ref{fig:m2}, where we train a single classifier that learns to optimize a joint loss function pertaining to the two tasks. 

We define a weight matrix $\mathbf{W}\in \mathbb{R}^{\vert Y \vert \times d}$ containing a set of weight vectors $\mathbf{w}_{\{y^s,y^e\}}$ for classifying each of the $\vert Y \vert$ possible label tuples $\{y^s,y^e\}$, where $\vert \cdot \vert $ denotes the cardinality of the set. In order to jointly model spontaneity and emotion, we intend to minimize a loss function $\mathcal{L}(\mathbf{W},Y,\mathbf{F})$ defined as follows. 
\begin{equation}
\label{eq:loss}
%\begin{aligned}
\mathcal{L}(\mathbf{W}, Y, \mathbf{F}) = \frac{1}{2}\sum_{(y^s, y^e) \in Y}\vert\vert \mathbf{w}_{\{y^s,y^e\}}  \vert\vert^2 + \mathbf{C}\sum_{j =1}^N\zeta_{j} \newline\\
%\end{aligned}
\end{equation}
The loss function $\mathcal{L}$ is sum of a a regularization loss term $\Vert \mathbf{w}_{\{y^s,y^e\}} \Vert$ and soft-margin loss term (optimization with slacks) i.e., $\zeta_j$. The parameter $\mathbf{C}$ controls the relative balance of the two cost terms. The term $\zeta_j$ allows for misclassification of the near-margin training samples while penalizing $\mathcal{L}$ by imposing a loss term that varies on the degree of the misclassification.
The optimal classifier weights $\mathbf{W}^*$ are then learned by minimizing the joint loss function $\mathcal{L}(\mathbf{W},Y,\mathbf{F})$ as 
\begin{equation}
\mathbf{W}^* \leftarrow \argmin_{\mathbf{W} \in \mathbb{R}^{\vert Y \vert \times d}} \mathcal{L}(\mathbf{W}, Y, \mathbf{F})
\end{equation}

\noindent The classifier is trained i.e., $\mathbf{W}^*$ is learned using the entire $\Omega_{train}$ using the same set of features described earlier. Since emotion can vary between two consecutive recordings, the joint model uses a sequence length of $\ell = 1$. 
%-------------------------------------
\section{Performance Evaluation}
\label{sec:typestyle}
We perform detailed experiments on the IEMOCAP database \cite{iemocap} to demonstrate the importance of spontaneity in the context of emotion recognition, and to validate the proposed classification models. 
\subsection{Experimental setup}
\begin{figure}[tb]
\centering
\includegraphics[width=0.9\linewidth]{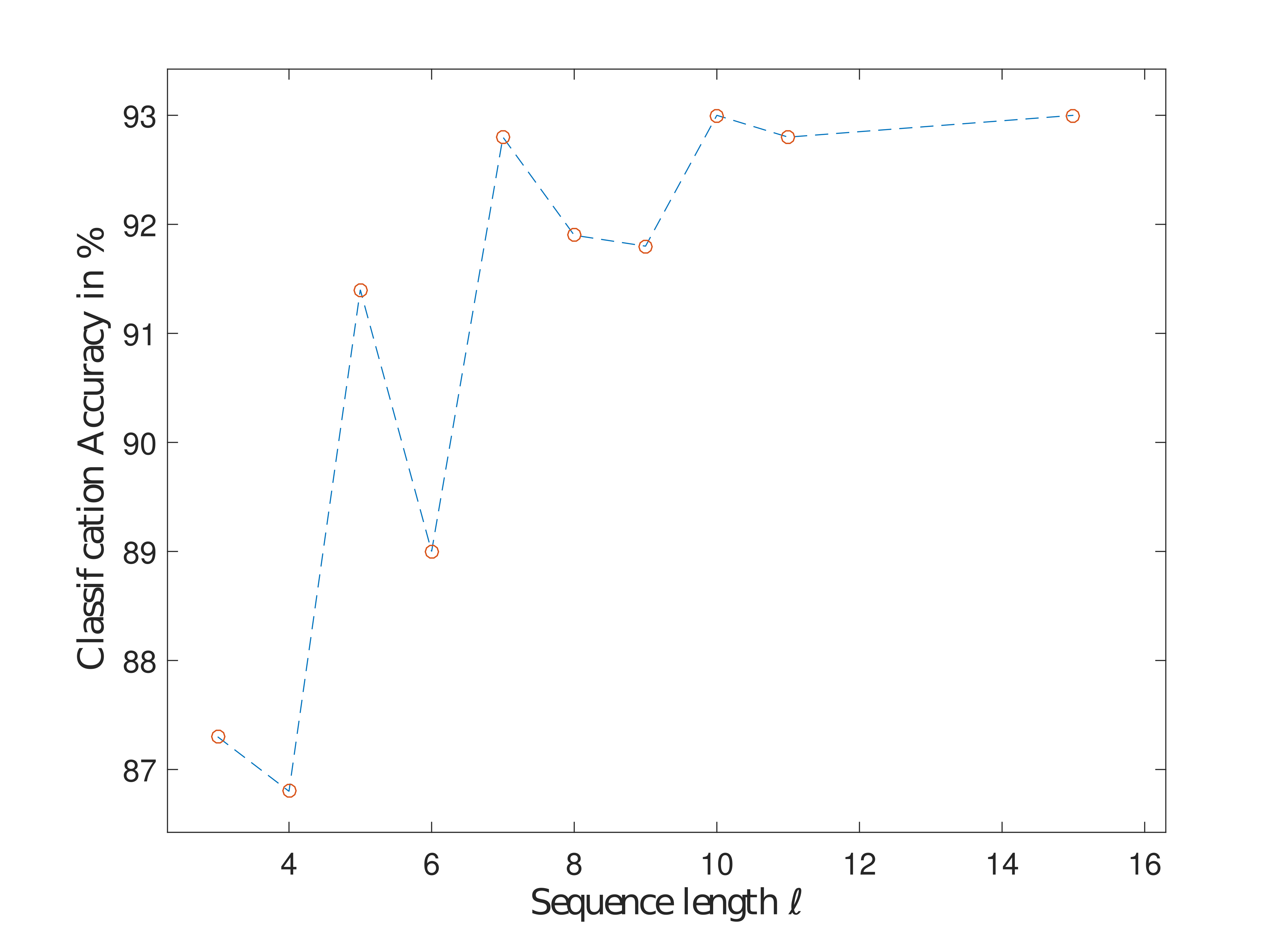}
\caption{Effect of varying context ($\ell$) on spontaneity classification.}
\label{fig:seqlen}
\end{figure}
\noindent\textbf{\emph{Database:}} We used the \emph{USC-IEMOCAP database} \cite{iemocap} for performance evaluation. It comprises $12$ hours of audiovisual data along with motion capture (mocap) recordings of face and text transcriptions. The data is collected in $5$ different sessions, and each session contains several dyadic conversations. Altogether there are $151$ conversations, which are labeled either improvised (spontaneous) or scripted. This serves as our spontaneity label $y^s$. There are almost equal number of scripted ($52.2\%$) and spontaneous ($47.8\%$) conversations in this database. Each conversation is further broken down to separate samples or utterances, which are organized speaker-wise in a turn-by-turn fashion. All samples are labeled by multiple annotators into one or more of the following six categories - neutral, joy, sadness, anger, frustration and excitement. A single sample may have multiple labels owing to different annotators. In such cases, the final label is chosen to be the label noted by the most annotators and randomly  between all the leading labels in case of a tie. We used the four emotion categories: \emph{anger}, \emph{joy}, \emph{neutral}, and \emph{sadness}.\\

\vspace{-0.015in}
\noindent\textbf{\emph{Parameter settings}:} The features described in Section \ref{subsec:features} are computed using a sliding window of length $w=25$ms with a stride of $m =10$ms. This yields a local feature vector of dimension $k=15$, and a global feature vector $\mathbf{f}$ of dimension $d=360$ for each sample. The features are normalized to have values between $-1$ to $1$. The SVMs use the radial basis function (RBF) kernel. All results reported are the average statistics computed over a $5$-fold cross validation. 

\subsection{Understanding spontaneity}
\label{subsec:impspontaneity}
\begin{table}[t]
\caption{Effect of features on spontaneity classification accuracy (in \%)} 
\centering
\begin{tabular}{c|cc}
\toprule
\textbf{Feature(s) removed} &  $\ell =  5$                  & $\ell = 10$                       \\ \hline 
None                            & $91.4$                   & $93.0$                   \\ 
ZCR                                    & $91.0$                   & $92.4$                   \\ 
VP                                     & $90.6$                   & $92.6$                   \\ 
F0                                     & $90.5$                   & $92.6$                   \\ 
MFCC                                   & $83.4$                   & $85.5$                   \\ 
VP, MFCC                              & $80.7$ &             $83.8$ \\
F0, MFCC                              & $83.2$ &             $84.9$ \\ 
ZCR, MFCC                             & $78.8$ &             $82.3$ \\ 
VP, F0                                & $90.6$                   & $91.5$                   \\ 
VP, ZCR                               & $90.2$                   & $92.1$                   \\
F0, ZCR                               & $90.6$                   & $92.2$                   \\
VP, ZCR, F0                          & $83.7$ & $91.9$ \\
Any two, MFCC                         & $<76$            & $<80$          \\ \hline
\end{tabular}

\label{featureexcl}
\end{table}
\begin{table}[t]
\caption{Effect of keeping only the delta features on spontaneity classification accuracy (in \%)} 
\centering
\begin{tabular}{c|cc}
\toprule

\textbf{Feature(s) removed} &  $\ell =  5$                  & $\ell = 10$                       \\ \hline 
None                            & $91.4$                   & $93.0$                   \\ 
ZCR                                  & $91.0$                   & $92.4$                       \\ 
VP                                    & $91.0$                   & $92.2$                   \\ 
F0                                   & $90.9$                   & $92.7$                       \\ 
MFCC                                  & $86.8$                   & $90.1$                   \\ 
 \hline
\end{tabular}

\label{stars}
\end{table}
 
\begin{table}[t]
\centering
\caption{Emotion recognition results for individual classes in terms of weighted accuracy (in \%).}
\def\arraystretch{1.5}%
\begin{tabular}{ c |c c c c}
\hline
 & \bf Anger & \bf Joy & \bf Neutral  & \bf Sadness \\ \hline \hline
 SVM baseline & $69.2$  & $37.0$  & $62.9$ & $\mathbf{76.9}$ \\
 RF baseline  & $73.1$& $6.1$  &$\mathbf{78.8}$   & $64.6$   \\ 
CNN-based \cite{neumann2017attentive}  & $58.2$ & $\mathbf{51.9}$ & $52.8$ &  $66.5$ \\ 
  Rep.\ learning \cite{ghosh2016representation}  &$53.5$ &$36.9$ &$52.6$ &  $64.3$ \\ \hline
 \multicolumn{5}{c}{Spontaneity-aware methods}\\ \hline
 \textbf{Hierarchical} & $\mathbf{80.2}$  & $37.5$  & $65.9$  & $73.3$  \\
 \bf Joint  & $71.2$& $13.1$& $75.9$ & $76.3$ \\ \hline
\end{tabular}
\label{tab:results1}
\end{table}
\begin{table}[t]
\centering
\caption{Emotion recognition results for all classes together in terms of weighted accuracy (in \%). }
\def\arraystretch{1.5}%
\begin{tabular}{ c | c c | c }
\hline
  & \bf Scripted & \bf Spontaneous & \bf Overall\\ \hline \hline
 SVM baseline & $56.8$ & $73.0$  & $65.4$\\
 RF baseline  & $62.1$ & $66.0$  &$64.1$  \\ 
 CNN-based \cite{neumann2017attentive}  &$53.2$ & $62.1$ & $56.1$  \\ 
 Rep.\ learning \cite{ghosh2016representation}  &- & $52.8$ & $50.4$   \\ \hline
 \multicolumn{4}{c}{Spontaneity-aware methods}\\ \hline
 LSTM \cite{emointerspeech} & -  & - & $56.7$  \\
 \textbf{Hierarchical}  & $\mathbf{64.2}$  &$\mathbf{74.0}$  & $\mathbf{69.1}$ \\
 \bf Joint &  $63.2$ & $69.8$ & $66.1$ \\ \hline
\end{tabular}
\label{tab:results2}
\end{table}
\noindent\textbf{\emph{Sanity check:}} We started with the hypothesis that emotional content is different in the spontaneous vs.\ scripted speech. In order to check this experimentally, we trained an SVM (under the same experimental conditions described in the previous section) on $\Omega_{train}$ of the IEMOCAP database that can discriminate among the anger, joy, neutral and sadness. During the test phase, we computed the recognition accuracy for the spontaneous and scripted speech separately. We observe that while the overall accuracy (using the baseline SVM) of emotion recognition is $65.4\%$, recognition accuracy is higher for speech samples labeled spontaneous i.e.\ $73.0\%$ and $56.8\%$ for scripted speech (see Table \ref{tab:results2}). This basic result supports our assumption that emotional content is different in spontaneous vs.\ scripted speech. This observation is also consistent with the results reported in a CNN-based recent work \cite{neumann2017attentive}.\\

\vspace{-0.02in}\noindent \textbf{\emph{Role of context in spontaneity:}} 
We performed spontaneity classification to understand the role of various parameters. We investigate the role of context and the contribution of various features in spontaneity detection. We train an SVM classifier with RBF kernel to distinguish between spontaneous and scripted speech using the features described in Section \ref{subsec:features}. At the utterance level, i.e.\ for $\ell=1$, this system achieves an average accuracy of $\mathbf{80\%}$. In order to study the effect of context on spontaneity classification, we vary the sequence length $\ell$. To account for longer context, we increase $\ell$ by concatenating consecutive utterances. Consequently, we concatenate their corresponding global features. This yields a feature vector $\mathbf{F}_\ell\in\mathbb{R}^{d\time \ell}$. The variation of classification accuracy with different values of $\ell$ is shown in Fig.~\ref{fig:seqlen}. The general trend is that the classification accuracy improves with the longer context (sequence length), and achieves $\mathbf{93\%}$ accuracy for $\ell = 10$. This result can be intuitively explained by the fact that as longer parts of the conversation is used for classification, it becomes easier to detect spontaneity. The result also highlights that spontaneity can be detected with a fairly high accuracy and hence assures us that an additional spontaneity detection module would not harm the overall performance of a speech processing pipleline because of incorrect detection of spontaneity.

\vspace{-0.02in}\noindent \textbf{\emph{Role of features:}} We investigate the importance of each feature individually in spontaneity classification by performing an ablation study. We exclude one or more of the LLD features at a time, and record the corresponding spontaneity classification accuracy. From the results presented in Table~\ref{featureexcl}, we observe that (i) MFCC features are the most important of all. (ii) Any single LLD feature can provide an accuracy of $\sim75\%$ indicating that any LLD feature is well suited for the task of spontaneity classification. Moreover, comparing the accuracies achieved when removing both the delta and the actual features (as in Table~\ref{featureexcl}) to removing the actual features but retaining the delta features (see Table~\ref{stars}), we notice that the delta features play a more crucial role than the original features themselves for the task of spontaneity classification.

\subsection{Emotion recognition results}
To compare the gain from using the spontaneity information, we construct two baselines: an SVM-based emotion classifier, and a random forest (RF)-based emotion classifier. Both classifiers are trained to recognize emotion without using any information about the spontaneity labels. We also compare with two other recent work on emotion recognition: CNN-based emotion recognition \cite{neumann2017attentive} and representation learning-based emotion recognition \cite{ghosh2016representation}. Additionally, we compare our results with a recent LSTM-based framework \cite{emointerspeech} that uses gender and spontaneity information for emotion classification.
The performances of the proposed spontaneity-aware emotion recognition methods (i.e., hierarchical and joint) along with those for the baselines and other existing methods are presented in Table \ref{tab:results1} and Table \ref{tab:results2}. The proposed hierarchical SVM outperforms the baselines and all other competing methods by achieving an overall recognition accuracy of $69.1$\%.  The joint SVM model achieves an accuracy of $ 66.1$\%. Comparing the performance of the baseline SVM with the proposed spontaneity aware SVM methods, we observe that even with the same features and classifier more than $3\%$ improvement for the hierarchical method in overall emotion recognition accuracy is achieved just by adding spontaneity information (see Table \ref{tab:results2}). Looking at the improvements in individual classes, the class \emph{anger} benefits the most from spontaneity information. This is evident from a notable $11\%$ improvement in recognition accuracy while using the hierarchical model over the SVM baseline (see Table \ref{tab:results1}. On the other hand, \emph{neutral} shows $3\%$ improvement, and \emph{joy} is only slightly affected by spontaneity as we compare the hierarchical method to the SVM baseline. \emph{Sadness} does not show any improvement when using spontaneity. The individual emotion recognition accuracies may possibly indicate that anger is a more spontaneous emotion (i.e., difficult to fake) than other emotions, such as sadness. Table \ref{tab:results2} shows that recognition accuracy is always lower for scripted speech irrespective of the classification method used. This indicates that emotion is easier to detect in spontaneous speech, and this result is consistent with the observations made in an earlier work \cite{neumann2017attentive}.

The proposed hierarchical classifier performs slightly better than our joint classifier possibly owing to the more accurate spontaneity classification. Recall that the spontaneity classifier for the hierarchical model used longer context ($\ell=10$) while the joint model uses $\ell=1$. Nevertheless, the joint classifier is still of practical use in the scenario when the temporal sequence of the recording is unknown, and hence the sequence length for spontaneity is necessarily constrained. 

Clearly, spontaneity information helps emotion recognition. Our SVM-based methods could achieve better result than all competing methods by explicitly detecting and using spontaneity  information in speech. The reason behind our SVM-based methods outperforming deep learning-based methods (e.g., CNN-based \cite{neumann2017attentive}, Rep. learning \cite{ghosh2016representation} ) is possibly the use of spontaneity and a longer context in the case of hierarchical method. The LSTM-based spontaneity aware method though uses the same four classes as ours, they use an aggregated corpus (using IEMOCAP and other databases) for training the LSTM network. Such training is different from our experimental setting. 
\section{Conclusion}
In this paper, we studied how spontaneity information in speech can inform and improve an emotion recognition system. The primary goal of this work is to study the aspects of data that can inform an emotion recognition system, and also to gain insights to the relationship between spontaneous speech and the task of emotion recognition. To this end, we investigated two supervised schemes that utilize spontaneity to improve emotion classification: a multilabel hierarchical model that performs spontaneity classification before emotion recognition, and a multitask learning model that jointly learns to classify both spontaneity and emotion. Through various experiments, we showed that spontaneity is a useful information for speech emotion recognition, and can significantly improve the recognition rate. Our method achieves state-of-the-art recognition accuracy (4-class, 69.1\%) on the IEMOCAP database. Future work could be directed towards understanding the effect of other meta information, such as age and gender.
\section{Acknowledgement}
The authors would like to thank SAIL, USC for providing access to the IEMOCAP database.
\balance
\bibliographystyle{IEEEtran}
\bibliography{ref}
\end{document}